# The Study of TVS Trigger Geometry and Triggered Vacuum Conditions

**Wung-Hoa Park, Moo-Sang Kim, Yoon-Kyoo Son, Byung-Joon Lee**

*Pohang Accelerator Laboratory, Pohang University of Science and Technology, Pohang 790-834*

**Klaus Frank**

*Department of Physics, Friedrich Alexander University, 91052 Erlangen Germany*

This presentation focuses on the optimization of the trigger unit of a six-rod TVS. The different configurations of the trigger pin and of the trigger electrode have been considered to study the electric field distribution at the triple points of the unit embedded in the cathode. To optimize the field enhancement, electric field simulations with a planar and a circular heads of the trigger pin in combinations with a convex and a concave shaped trigger electrodes have been done. The simulations were done with an applied trigger pulse voltage of $U_{trigger}$ = 5 kV and with a discharge voltage the main switch of $U_{switch}$ = 20 kV. The experimental values had been $U_{trigger}$ = 40 kV and $U_{switch}$ = 5 kV. The simulation results show that the combination of a circular trigger pin head and a concave trigger electrode yields the highest electric field of $9.6 \cdot 10^6$ V/m at the triple point. In-parallel experiments have been performed with those four trigger configurations. The results of the experiments however cannot yet clearly confirm the trend in the results of the field simulations.

Email: whpark92@postech.ac.kr

Fax: +82-54-279-1399

## I. INTRODUCTION

Triggered Vacuum Switches (TVS) with the electrode system as an array of alternative polarity rods spatially distributed around a circle have found a variety of applications in pulsed power technology. Usually switching-on a TVS is carried out with a trigger unit, which provides an appropriate voltage pulse on the triggering electrode with optimized parameters of the trigger current. This voltage pulse causes usually breakdown by a surface flashover on a ceramic disc of the triggering gap. This discharge current injects a plasma into the void of the vacuum gap in order to close it [1]. This result is an arc discharge in the vacuum gap between the main electrodes. As the requirements for very high peak currents and high charge transfer, the primary parallel-electrode configuration of the TVS was replaced by a multiple-rod configuration [2-5]. This new lay-out made it necessary to develop a special trigger structure with trigger pin and trigger electrode with a ceramic for surface flashover. As a consequence the degradation of the trigger unit has to be minimized with regard to its design and to its position in the main electrode. As the trigger surface is set in the center of main cathode, the discharge of main gap is bound to influence the trigger surface, depending on its location there. The main discharge current accompanies metal vapor, ions, and electrons. The metal vapor is deposited on the insulator surface that reduces the trigger voltage. Additionally the trigger pin will be eroded by the high main current.

In this paper a TVS with six-rod gap is fabricated of OHFC copper, and four special trigger units with ceramic surface have been designed, it is shown in Fig. 1. In order to find optimized parameters for reliable triggering, the two different geometries of the trigger pin, one with a planar head and another with a circular head, have been chosen with the two different geometries of the trigger electrode, a convex and a concave type. For technical reasons the experiment was done with different parameters of trigger and switch voltage compared to that ones in the simulations. In order to guarantee reliable triggering the applied trigger voltage amplitude was selected to $U_{trigger}$ = 40 kV. The applied trigger pulse of four "virgin" trigger units has shown an initial breakdown voltage of 18 kV and 15 kV, respectively. By repetitive pulsing, after 250 shots all breakdown voltages converged to an average value of < $U_{trigger}$ > = 5 kV, which was in the simulations of all four configurations later used (see Section II). It is assumed, that the obtained results however are independent of the differing parameters.

## II. EXPERIMENTAL Set-Up and Trigger Configurations

- Experimental Set-Up

The lay-out of the experiment is shown in Fig. 2 and in Table 1 the most important parameters are listed. The electric circuit comprehends a capacitor bank (6 capacitors in parallel) with a total capacitance of 16.63 μF generating a peak current of 4.5 kA at an operating voltage of U $_{switch}$ = 5 kV. The total inductance is 18.7 μH. The total energy stored, is 3.3 kJ. In addition, a power supply, the trigger pulse generator and a vacuum pumping unit complete the set-up. The TVS was installed in an enclosed chamber with its vacuum pressure in the range of $10^{-6} - 10^{-4}$ mbar. Different probes have been installed in the set-up to monitor and record the data of the discharge voltage, the current and the time constants.

- Trigger Configurations

Basically the trigger device of the TVS consists a trigger pin and an insulator which are inserted in the main electrode, in the cathode [5]. As the parameters of the trigger pulse like current rise time, voltage, current, and waveform essentially determine the breakdown behavior of the main gap the trigger pulse generator is of decisive importance, but in this paper the characteristics of the trigger pulse generator are not discussed in detail. The emphasis was on investigating basic parameters of the trigger unit, namely geometry of the trigger pin and the trigger electrode. Both influence the electric field distribution at the triple points of the trigger, inserted into the cathode.

With the design of the trigger units the goal was to find a geometry with an optimum electric field strength at the triple points. The material of trigger pin is tungsten. The ceramic plate of the trigger units are made of $Al_2O_3$. The trigger pulse generator delivers a sinusoidal voltage pulse with $U_{trigger}$ = 40 kV applied to the trigger pin. The whole experimental set-up was on room temperature.

**III. ELECTRICAL FIELD SIMULATION OF THE FOUR DIFFERENT TRIGGER UNITS**

As described in the previous section the two trigger pin and two trigger electrode geometries allow to study four combinations of them: Those are:

- a planar trigger pin – a convex trigger electrode
- a planar trigger pin – a concave trigger electrode
- a circular trigger pin – a convex trigger electrode
- a circular trigger pin – a concave trigger electrode

These combinations have been simulated with the commercial computer program CST [6], which is widely used in electric field simulation. The results are shown in Fig. 3 - 6. In these figures, the upper graph shows the spatial distribution of the electric field strength and the lower graph is the strength of the electric as function of distance, where the direction of distance goes from the left of the triple point. From the simulations, the average trigger voltage, $< U_{trigger} >$ = 5 kV corresponding to the main electrode voltage is $U_{switch}$ = 20 kV was taken (see Section I). With these parameters the maximum values of electric field strength is $7.2 \times 10^6$ V/m for the planar-convex, $7.5 \times 10^6$ V/m for the circular-convex, $7.6 \times 10^6$ V/m for concave-planar and $9.6 \times 10^6$ V/m for the concave-circular, respectively. This result shows that the combination of a circular trigger pin – a concave trigger electrode generates the highest electric field at the triple point compared to other three combinations.

## IV. Experimental Results

With the experimental set-up (see Fig. 2) the following signals have been recorded: The voltage signal of trigger pulse synchronized with the monitor output signal of the trigger pulse generator and current and voltage pulse of the main discharge, it is shown in Fig. 7. The main voltage is 5 kV, the maximum current is 22 kA and trigger voltage is 5 kV. The Breakdown voltage amplitude was measured up to 300 shots for every trigger configuration. The initial trigger breakdown voltage was 13-18 kV for the circular convex configuration as shown in Fig. 8. The trigger breakdown voltage was reduced to 3-4 kV after 300 shots. For all four trigger units there was similar behavior observed. The first 200 shots, the signals are superimposed by strong fluctuations in amplitude. The causes for these fluctuations are certainly not noise from the main discharge. However, the reason it unclear, it is considered by the metal deposition in the trigger gap, the erosion by the high current, and so on. For the convex type of the trigger electrode the fluctuations are larger compared to those of the concave type. For 250 discharges and more, the fluctuations are getting less accompanied with an asymptotic decrease of the trigger breakdown voltages to an average value of $<U_{trigger}> = 5$ kV, indicating a kind of "conditioning". Unfortunately the experimental results do not confirm the results of the field simulations, namely that the circular pin – concave trigger electrode configuration should have the lowest breakdown voltage. To understand these contradictory results further experiments have to be done, especially studying the important parameters like delay and jitter of trigger breakdown.

## V. CONCLUSIONS

In this paper, it was observed by field simulations that the circular trigger pin - concave trigger electrode configurations yields the strongest electrical field at the triple point. However, there was no explicit evidence from experiments to find the best trigger results from any of the configuration. And the relation of the electric field distribution on triple points of four trigger pin with simulation to the trigger breakdown voltage for electrode configurations is unclear. In next experiments, to find a proper condition of the unit, the breakdown voltage of the trigger will be measured without the main discharge and a delay and a jitter will be also studied. To conclude: The electric field simulation is a valuable tool to optimize and study the variable geometries of the TVS trigger. However, the information of parameters considered in the simulation program is not enough to explain the results of experiments, to research it, the analysis of the ceramic insulator will be accomplished with the study of the proper trigger conditioning.

**ACKNOWLEDGEMENT**

This project was supported by the Agency for Defense Development (ADD) under grant number 14-BR-EN-22. In addition, the authors would like to thank all members of the MKV who assisted and supported with their advice the installation of the experimental set-up. Special thanks to Dong Hyun Kim, who helped with technical knowhow generally, and for the vacuum system in particular.

Characteristics of a Surface Breakdown Triggered Vacuum Switch with Six Gap Rod Electrode System", *IEEE Transactions on Dielectrics and Electrical Insulation Vol. 18(4), pp.997 – 1002.*

[6] https://www.cst.com/.

Table 1. Parameters of Experimental Set-Up.

| Main Voltage (simulation/experiment) | 20 kV / 5 kV |
|---|---|
| Current (simulation/experiment) | 17.8 kA / 4.5 kA |
| Total capacitance | 16.63 µF |
| Reactance | 18.7 µH |
| System impedance | 1.12 Ω |
| Trigger pulse (experiment/simulation) | 40 kV applied / 5 kV average |
| Total energy (simulation/experiment) | 3.3 kJ (0.2 kJ) |
| Trigger pin | tungsten |
| Trigger electrode | copper |
| Main electrodes (anode, cathode) | copper |
| Vacuum | $10^{-6} - 10^{-4}$ mbar |

Figure Captions.

Fig. 1. The TVS electrodes, trigger electrodes, and trigger pins.

Fig. 2. The electrical TVS test system.

Fig. 3. The electric field of planar trigger pin and convex trigger electrode.

Fig. 4. The electric field of circular trigger pin and convex trigger electrode.

Fig. 5. The electric field of planar trigger pin and concave trigger electrode.

Fig. 6. The electric field of circular trigger pin and concave trigger electrode.

Fig. 7. Oscilloscope signals (yellow: the TTL signal which is control signal of the trigger pulse generator, magenta: trigger voltage, sky-blue: main electrode voltage, green: main electrode current).

Fig. 8. The trigger voltage for number of trigger shots.

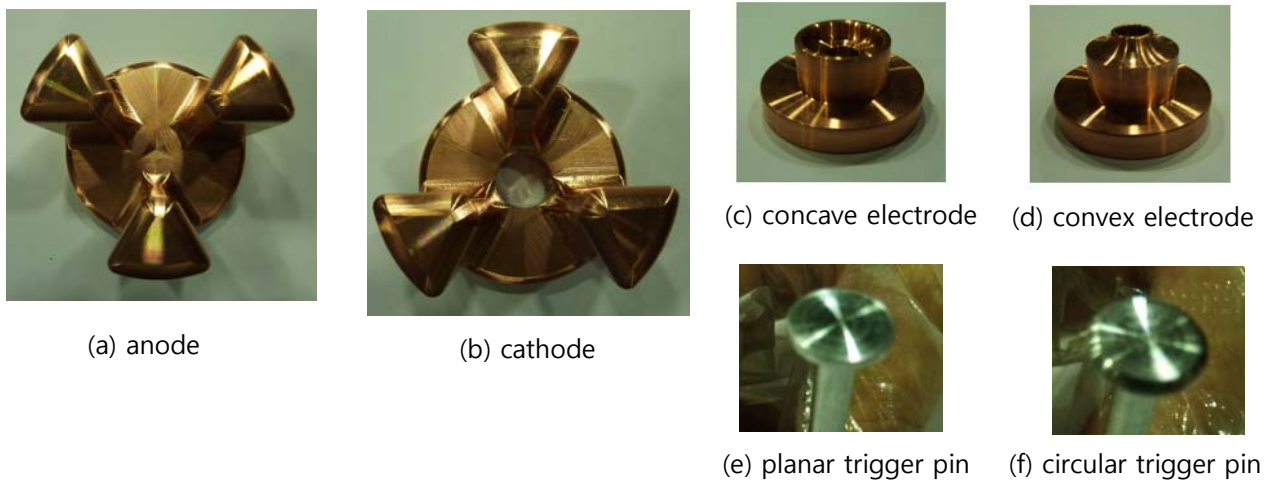

(a) anode  (b) cathode  (c) concave electrode  (d) convex electrode

(e) planar trigger pin  (f) circular trigger pin

Fig. 1.

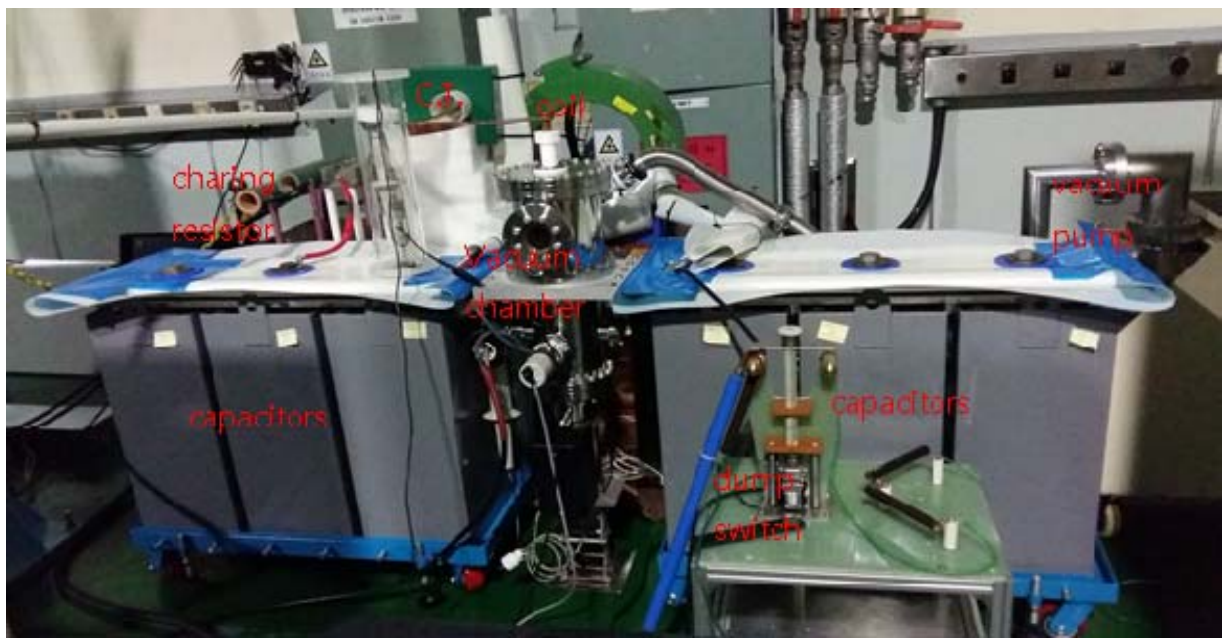

Fig. 2.

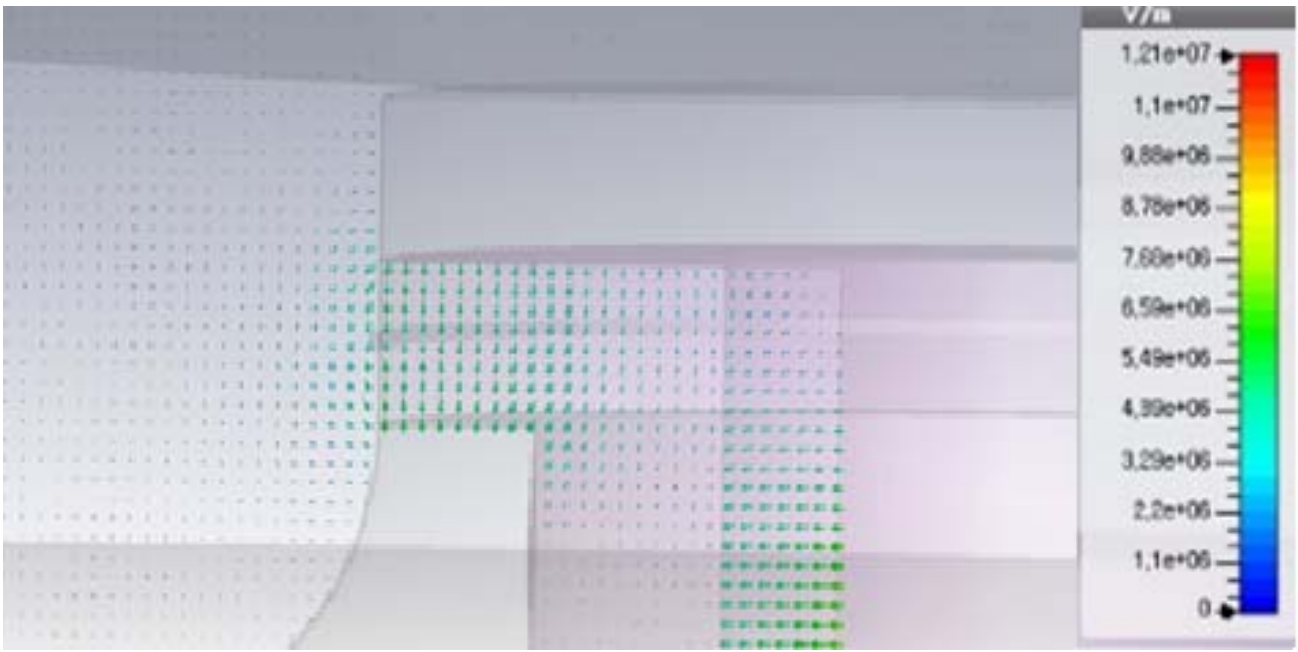

(a) The electric field of planar-convex (planar trigger pin and convex trigger electrode).

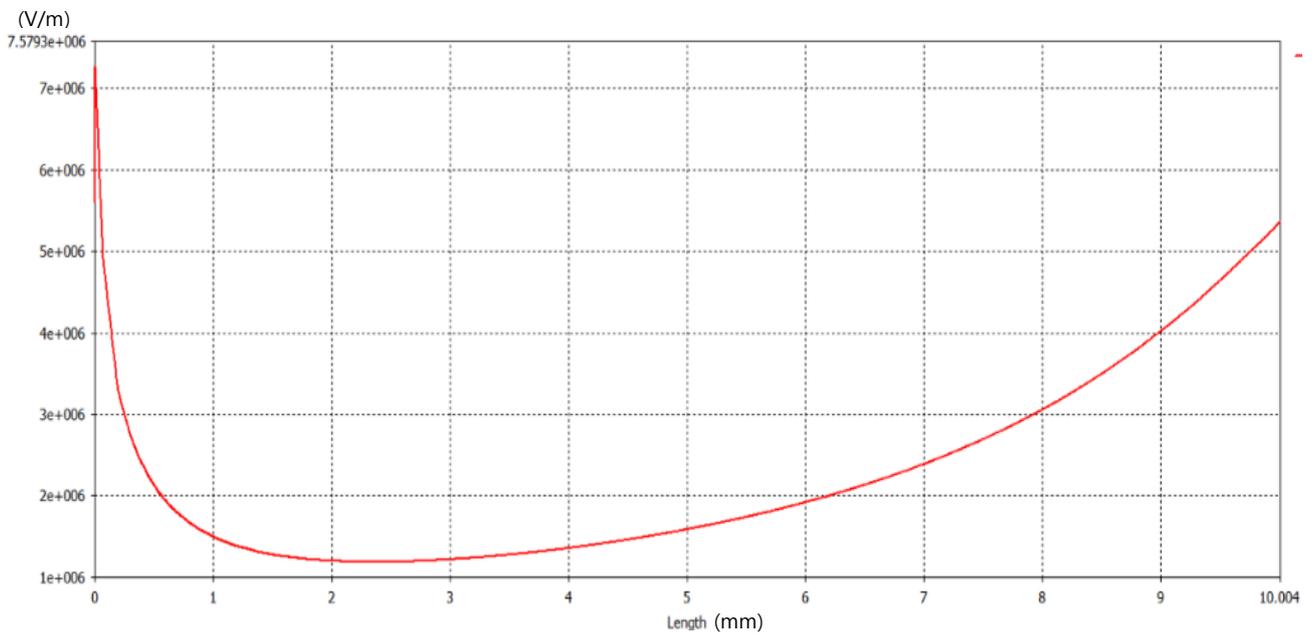

(b) The electric field strength for distance of the planar-convex.

Fig. 3.

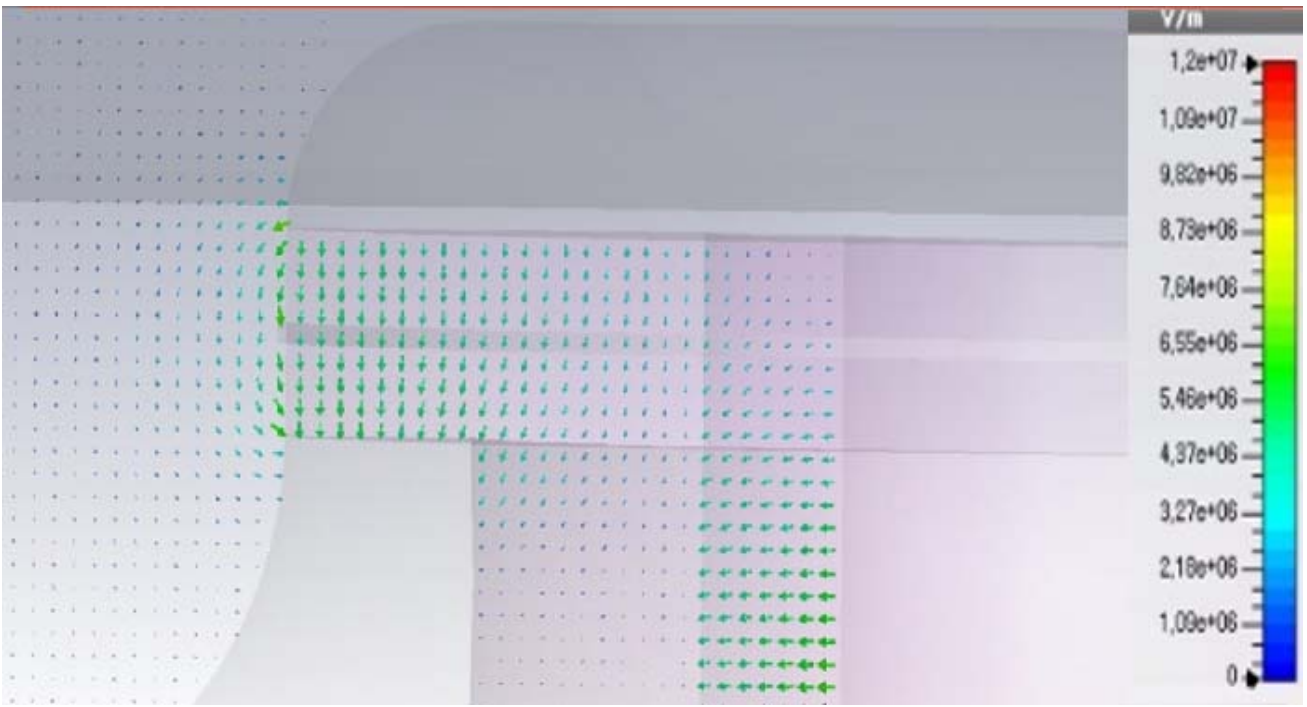

(a) The electric field of planar-convex (circular trigger pin and convex trigger electrode).

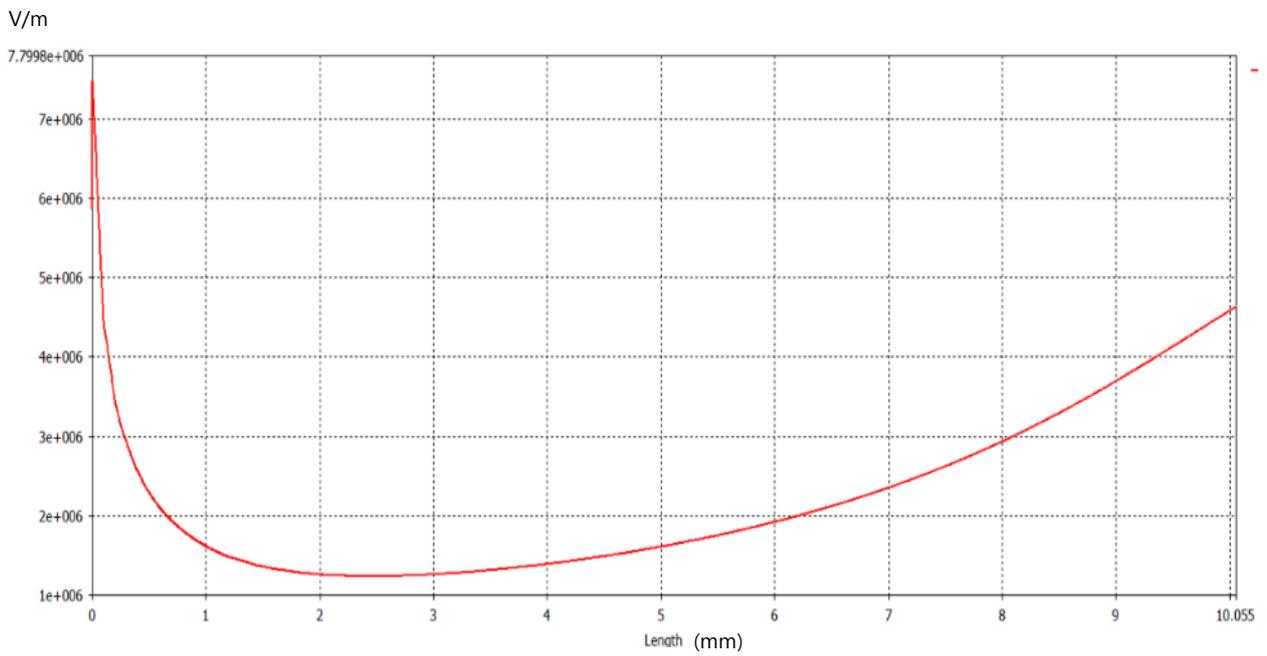

(b) The electric field strength for distance of the circular-convex.

Fig. 4

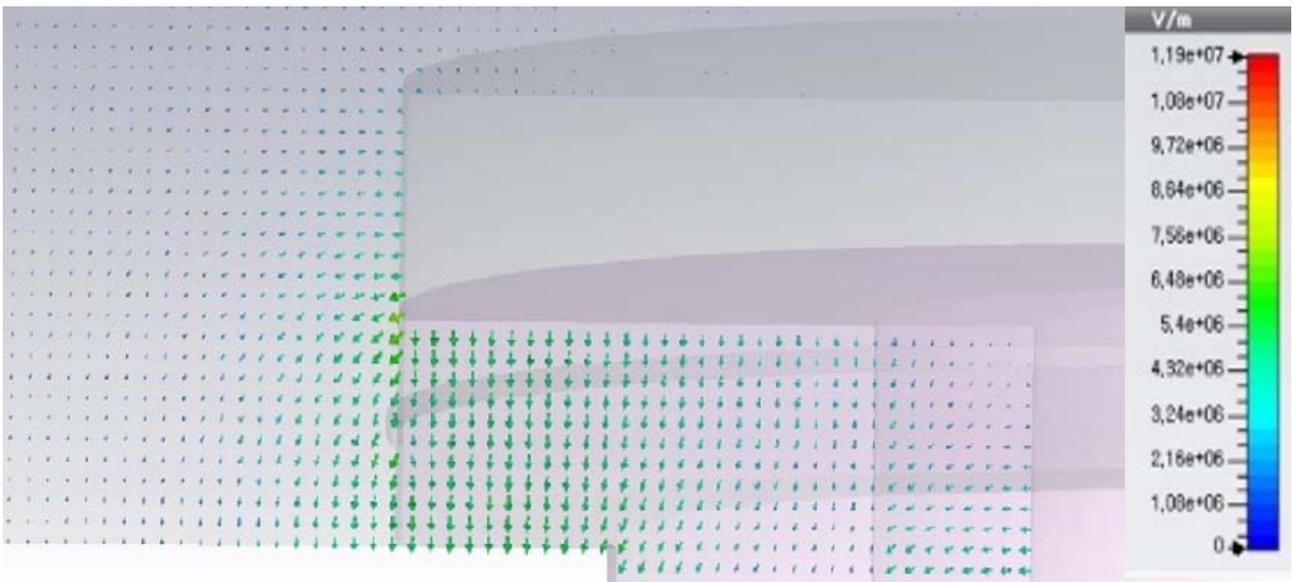

(a) The electric field of planar-concave (planar trigger pin and concave trigger electrode).

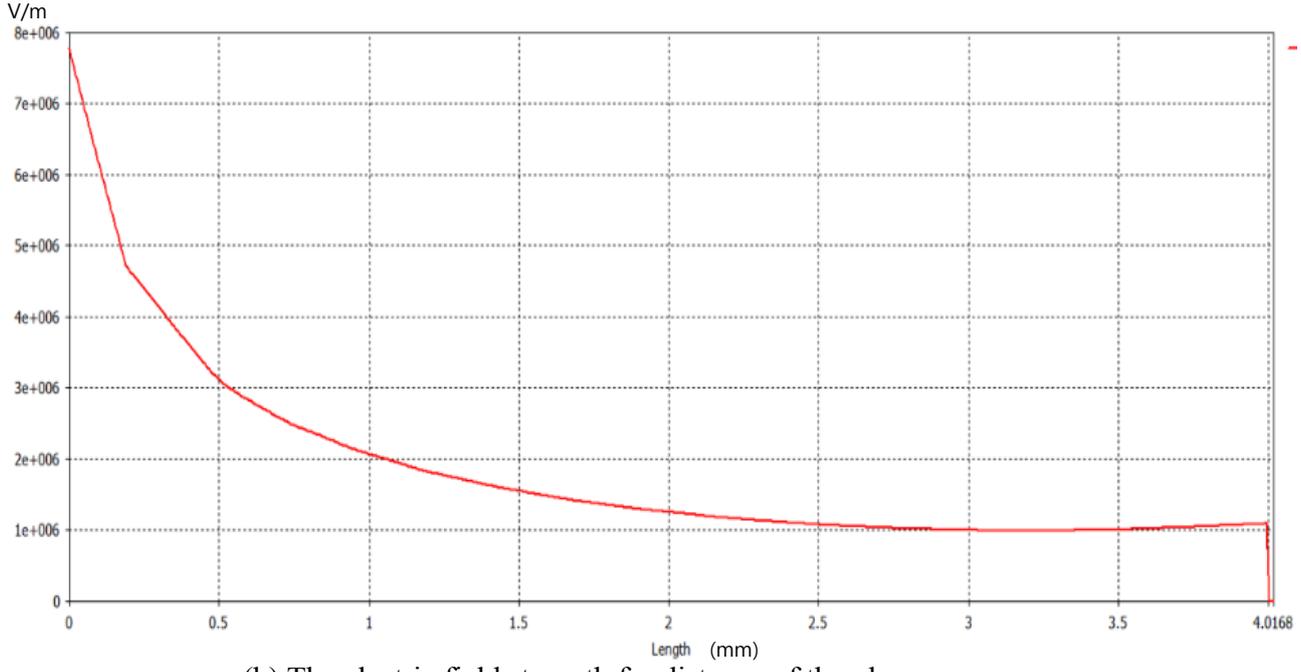

(b) The electric field strength for distance of the planar-concave.

Fig. 5

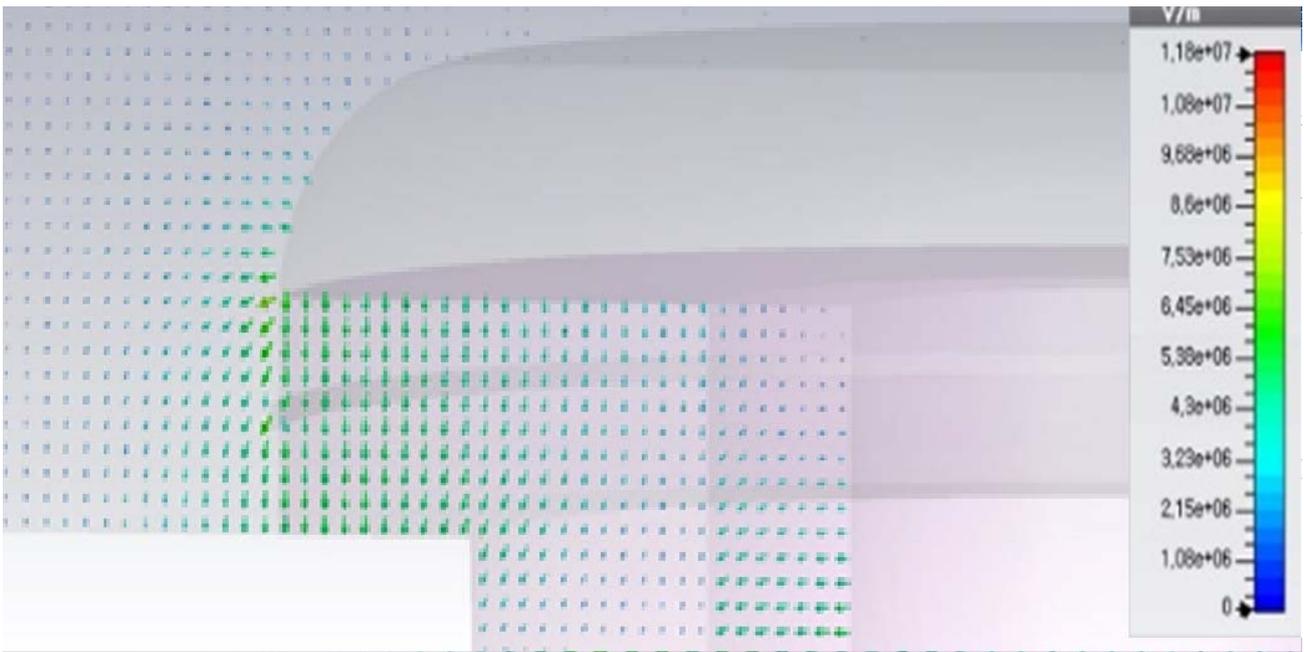

(a) The electric field of circular-concave (circular trigger pin and concave trigger electrode).

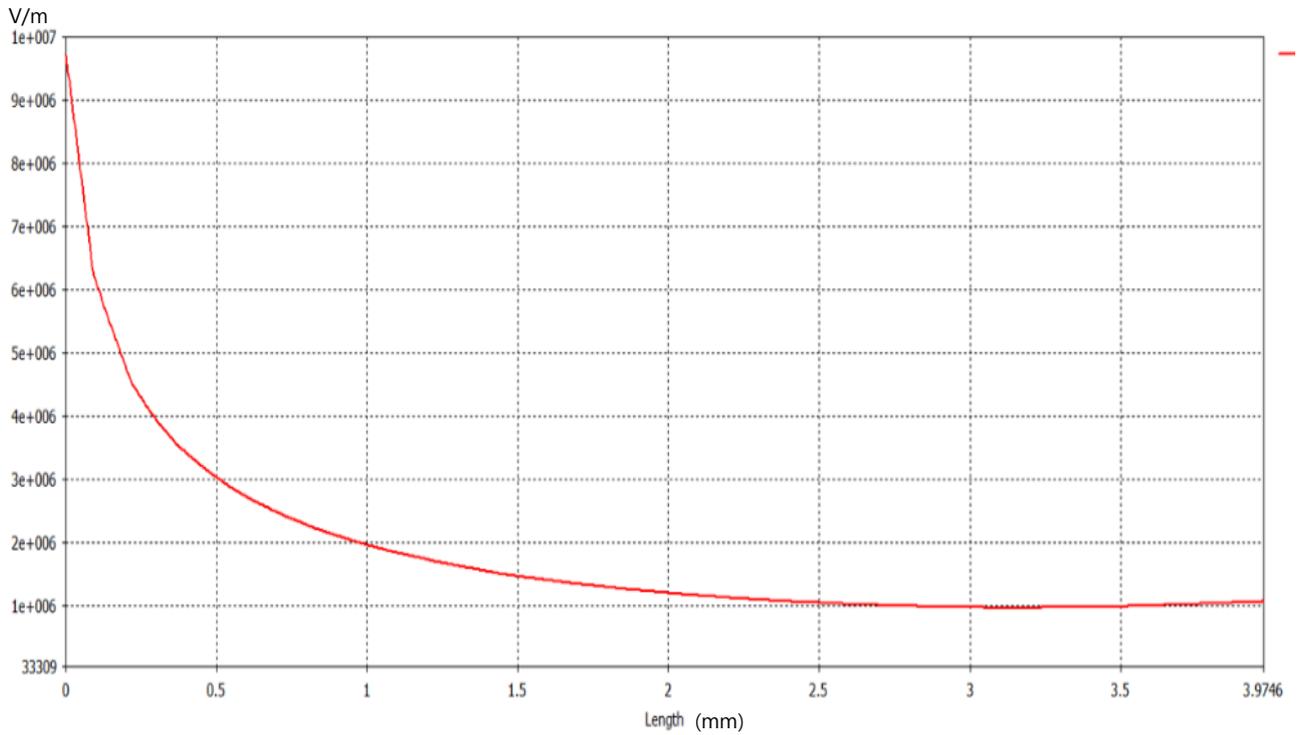

(b) The electric field strength for distance of the circular-concave.

Fig. 6

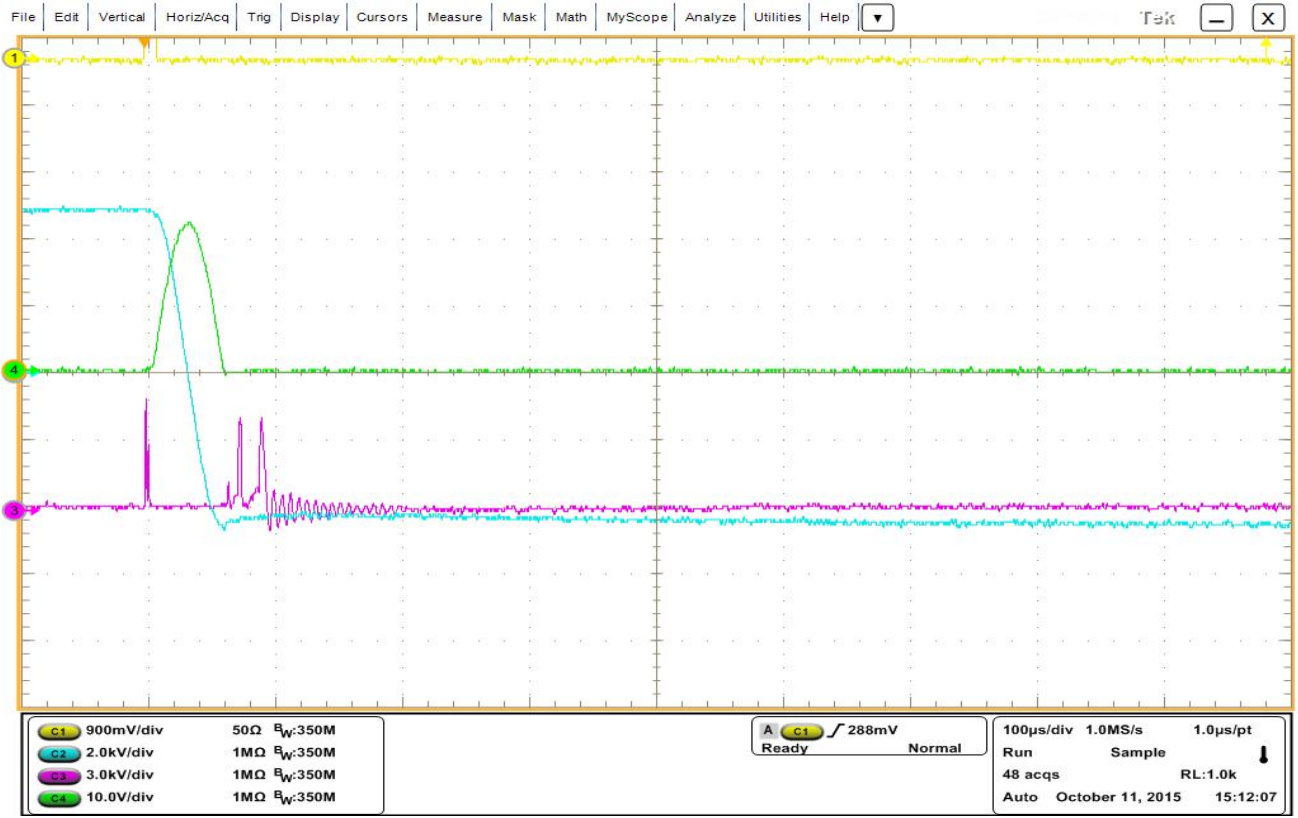

Fig. 7

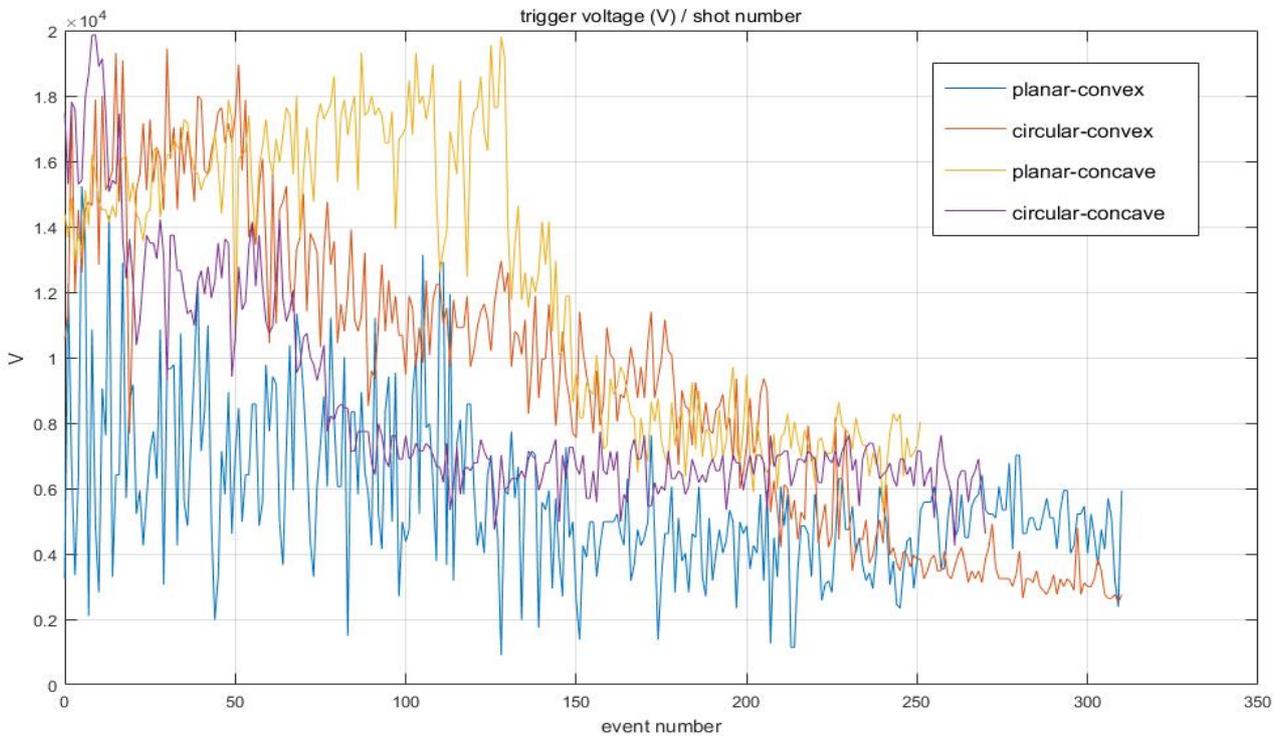

Fig. 8.

15